\documentclass{article}

\usepackage{arxiv}

\usepackage{floatrow}
\usepackage[utf8]{inputenc} % allow utf-8 input
\usepackage[T1]{fontenc}    % use 8-bit T1 fonts
\usepackage{hyperref}       % hyperlinks
\usepackage{url}            % simple URL typesetting
\usepackage{booktabs}       % professional-quality tables
\usepackage{amsfonts}       % blackboard math symbols
\usepackage{nicefrac}       % compact symbols for 1/2, etc.
\usepackage{microtype}      % microtypography
\usepackage{lipsum}
\usepackage{natbib}
\usepackage{graphicx}
\usepackage{floatrow}
\usepackage{amsmath}
\usepackage{amsfonts}
\usepackage{amssymb}

\newfloatcommand{capbtabbox}{table}[][\FBwidth]

\title{Data-driven modelling and characterisation of task completion sequences in online courses }

\author{
  Robert L.~Peach\\
  Department of Mathematics\\
  Imperial College London\\
  London, UK \\
  \texttt{r.peach13@imperial.ac.uk} \\
   \And
 Sam F.~Greenbury \\
  Department of Mathematics\\
  Imperial College London\\
  London, UK \\
   \And
 Iain G.~Johnston \\
  School of Biosciences\\
  University of Birmingham\\
  Birmingham, UK \\
     \And
 Sophia N.~Yaliraki \\
  Department of Chemistry\\
  Imperial College London\\
  London, UK \\
       \And
 David Lefevre \\
  Imperial Business School\\
  Imperial College London\\
  London, UK \\
       \And
  Mauricio Barahona\\
  Department of Mathematics\\
  Imperial College London\\
  London, UK \\
}

\begin{document}
\maketitle

\begin{abstract}

The intrinsic temporality of learning demands the adoption of methodologies capable of exploiting time-series information. 
In this study we leverage the sequence data framework and show how data-driven analysis of temporal sequences of task completion in online courses
%together with visualisation techniques,
can be used to characterise personal and group learners' behaviors, and to identify critical tasks and course sessions in a given course design. 
We also introduce a recently developed probabilistic Bayesian model to learn sequence trajectories of students and predict student performance.
The application of our data-driven sequence-based analyses to data from learners undertaking an on-line Business Management course reveals distinct behaviors within the cohort of learners, identifying learners or groups of learners that deviate from the nominal order expected in the course. 
Using course grades \textit{a posteriori}, we explore differences in behavior between high and low performing learners. We find that high performing learners follow the progression between weekly sessions more regularly than low performing learners, yet within each weekly session high performing learners are less tied to the nominal task order. We then model the sequences of high and low performance students using the probablistic Bayesian model and show that we can learn engagement behaviors associated with performance.
We also show that the data sequence framework can be used for task centric analysis; we identify critical junctures and differences among types of tasks within the course design. 
We find that non-rote learning tasks, such as interactive tasks or discussion posts, are correlated with higher performance. We discuss the application of such analytical techniques as an aid to course design, intervention, and student supervision.
\end{abstract}

% keywords can be removed

\section{Introduction}

Learning is a process that occurs over time~\citep{Wise2017}: new knowledge is built upon existing knowledge, suggesting that we should incorporate a temporal dimension in the analysis of learning data. 
%It is, therefore, important to capture the temporal relationships between events in a learning environment. 
Here, we develop a framework to model and analyse temporal sequences of task completion and task-to-task transitions by students taking a course, and allows us to compare them to the expected (i.e., designed) task order in the course. Our work introduces novel data-driven metrics and a Bayesian probabilistic model for analysis and prediction of sequence data, and shows that the order of task completion at fine-grained resolution facilitates improved prediction of performance and can be used to inform task-level course design. Unlike methods that construct feature vectors, our temporal data sequence framework~\citep{Mahzoon2018} directly incorporates temporal relationships between tasks. 
%Moreover, we claim that using a fine-grained sequential data model enables learning designers to take a critical view of their course design and make appropriate re-designs.
In the paper, we present the mathematical formulation and structure of the temporal data sequence framework, and we show how the data-driven analyses and Bayesian model can be used to predict student success, to reveal student confidence at a task level, and, from a task-centric perspective, to aid with learning re-design. We illustrate our work through data of student task completion collected from an online course taken over one semester as part of a 2-year MBA programme.

The availability of time-stamped data has improved substantially with the advent of computer-based education 
platforms~\citep{Kuzilek2017}, coupled with advances in educational data mining and learning analytics that 
provide new and innovative tools for automated analysis. 
However, despite this increase in available tools, temporal analyses are still understudied in education 
research~\citep{Wise2017,Barbera2015}.
%In addition to computer- based data, other sources of temporal learning data are also emerging \citep{Schneider2013,Andrade2017,Liu2018}. 
It is therefore important to expand the toolbox of temporal methodologies that are constructed to extract actionable information directly from the observed data without over-simplifying the learning process. For instance, the superiority of distributed learning over massed learning inherently posits the benefits of a consistent learning behaviour over time instead of an irregular, syncopated approach to learning~\citep{Bloom1981}. However, the distinction into only two such broad behaviours is an over-simplification, and a number of subtly different temporal behaviours have been shown to exist in real data~\citep{Peach2019}. In general, the wide availability of data opens the possibility to introduce unsupervised methodologies that do not reduce observations to predetermined, broad categories, and aim to extract more nuanced results and conclusions.

\subsection{Temporal feature analysis}

In educational data mining, the most common approach to temporal analysis is to describe each student through a \textit{feature vector} composed of a small selection of aggregated features from time-series data (e.g., time-gap between sessions, participation time, number of sessions, 
post views)~\citep{Kapur2008,Cepeda2006,Wise2012}. These feature vectors can be combined with additional non-temporal information, such as grades or demographic information. The position of each vector (i.e., each student) in feature space can be further analysed using supervised or unsupervised learning algorithms to make predictions or to investigate the structure of the data \citep{Munch2017}. For instance, \citep{Kloft2015} extracted features aggregated over the course of a week  (video views and active learning days combined with features such as country of origin) to predict drop-out rates. Similarly, \cite{Halawa2014} selected four predictive features (video-skipping, assignment skipping, lag, and assignment performance) to predict drop out rates. Feature analyses have also been highly successful at predicting stop out one-week in advance~\citep{Taylor2014}, yet the analysis of posting patterns of students on a forum over the course of a semester. 
%They explored the rhythm of students posting on a forum and found that students were posting more often during the weekdays than at the weekend. However, this approach 
did not reveal anything about learning outcomes and student engagement with forum threads~\citep{Haythornthwaite2012}.
Recently, unsupervised clustering of features from learners following a blended course on two platforms (face-to-face, online)~\citep{Carroll2017} identified four behavioural groups according to their differing levels of engagement across the two platforms. Other examples of temporal feature extraction include \cite{Aguiar2014}, who calculated the number of times a student logged into their account among other non-temporal features; \cite{Papamitsiou2014}, who extracted temporal trace data (e.g., time to answer correctly) to predict student performance; and \cite{Riel2018}, who extracted the time to half-way completion and frequency of completion, to develop tools for course managers or learning designers to understand and visualise participant performance.

Whilst feature vector analyses can be predictive of drop-out or performance in particular scenarios, these approaches are limited by the choice of features (`feature engineering'), itself limited by the granularity of the data. Indeed, extending the feature space to include higher resolution temporal features (e.g, timestamps of lecture viewings or start-dates of peer-graded assessments) has been shown to improve prediction accuracies for drop-out and final performance~\citep{Ye2014,Ye2015}. Despite such improvements, temporal feature vectors lead to a `temporally averaged' analysis of student trajectories, as they do not capture longitudinal (dynamical) changes, and therefore may miss important events or temporal dependencies.

% Temporal feature vectors requires \textit{a priori} selection of features to mine from the data. Such selection inherently biases the analysis to particular properties whilst removing potentially interesting and predictive information. 

\subsection{Theoretically-motivated temporal analyses}

Beyond feature vector representations, there have been developments guided by theoretical considerations~\citep{Wise2015}. For example, a temporal analysis of conversational data used segmentation (or `stanzas') to construct a network of micro-scale elements to reveal connections across timescales \citep{Lund2017}. A study of communal knowledge in online knowledge building discourse used temporal analytics combined with graph theory to identify ideas and their mobility over time~\citep{Vwen2017}. Graph centrality has been used to explore temporal and contextual profiles of shared epistemic agency on discourse transcripts~\citep{Oshima2018}, identifying pivotal points of discourse exchanges.
% however, were not yet able to understand what those pivotal points in time represented. 
\cite{Halatchliyski2014} emphasised the importance of temporality as the main component of learning analytics through main path analysis of Wikiversity domains to  model the flow of ideas. The method they developed could be used to support teachers in coordinating knowledge building, yet the method is not suited to sequential completion of tasks and would instead be interesting when applied to courses with no distinct task order. Another approach of temporal analysis is to perform so-called micro-analyses, where singular `extreme' learners are examined in depth to explore differing behaviours \citep{Wise2012}.  Whilst such approaches provide insight into particular behaviours, the conclusions are difficult to generalise and justify statistically \citep{Reimann2009}.

Given the recent renaissance of deep learning, there is opportunity to apply recurrent neural networks to temporal data; however, a recent effort by \cite{Tang2016} found that the granularity at which student action data was fed into the model did not provide consistently predictive results. Moreover, the insight of such analyses is often lost within the highly complex and non-linear prediction model \citep{Schmidhuber2015}.

Whilst each temporal analysis requires a theoretically justified model, there is clearly also a need for high-resolution data that can reveal nuanced longitudinal relationships. Considering this, \cite{Liu2018} used a coarse grained approach across knowledge components to identify focal points that were worthy of more in-depth analysis using high-resolution data and analysis. This method integrated the temporal and data resolution scales that often evade temporal analyses, however, this approach may ignore potentially important focal points that can only be identified using high resolution analysis from the start. \cite{Peach2019} used dynamic time warping on high-resolution task data to calculate similarities between the raw time-series of a cohort of students undertaking an on-line degree, and implemented a quasi-hierarchical clustering algorithm to identify data-driven groupings of students that exhibited similar temporal behaviours. They behavioural clusters were shown to be more predictive of final performance than classifiers based on temporal feature extraction method.
%Temporal analyses have also been used to identify focal points in a student learning trajectory. 

\subsection{Sequence based methods}

More recently, sequence-based methods have been shown to provide meaningful insights. For example \cite{Andrade2017} used optimal sequence matching to analyse the sequence of hand movements in a multi-modal learning environment.  They found that incorporating temporality into their analysis was crucial to find a correlation between sensorimotor coordination and learning gains. Another sequence-based method was implemented by \cite{Chen2017a}, who used Lag-sequential analysis (LSA) and Frequent Sequence Mining (FSM) for the purpose of discovering sequential patterns in collaborative learning. They found that both LSA and FSM were able to uncover productive threads for collaborative learning. Data sequence models were introduced by \cite{Mahzoon2018} to model within- and between-semester temporal patterns. Their model emphasised the importance of incorporating temporal relationships of heterogeneous data to more accurately predict success or failure of students relative to non-temporal models with the same data.  Sequence analysis has also been used on learning processes to test student learning outcomes at multiple levels and timescales \citep{chiu2005new,chiu2016statistical,chiu2018statistically}. \cite{chiu2018statistically} explored whether one or more temporal sequences of learning processes related to a later learning outcome relative to a random concatenation of learning processes.

% In particular, sequence-based methods have been shown to provide meaningful insight. For example, \cite{Andrade2017} used optimal sequence matching to analyze the sequence of hand movements in a multi-modal learning environment.

\subsection{Our contribution}

%The sequence data models defined by \cite{Mahzoon2018} constructs a series of nodes that each represent a period of time. Information about each student, such as courses taken, grades received, extra-curricular activities, is extracted and attributed to each temporal node. 

The recent success of sequence data analyses, as described above, suggests that using a sequence data framework can provide insightful analyses into learning analytics. We also highlight that the temporal sequence data framework is flexible: it provides an opportunity to analyse student trajectories as well as course learning design. A study by \cite{Mendez2014} concluded that using analytical techniques, including temporal analyses such as sequence mining, could highlight interesting anomalies or irregularities that can be the object of a more focused investigation by individuals with deep knowledge of the course. Accordingly, we use the data sequence framework to take a task-centric perspective, and identify common irregularities in task sequences across students.

%necessary to pursue and further extend this framework.

Despite the benefits of sequence data models, there is a need for sequence-based methods for the analysis of  high-resolution temporal data\citep{Ye2014,Ye2015,Tang2016,Mendez2014,Liu2018}. In this paper, we treat each node in our sequence data framework as a single event allowing us to analyse the ordering of tasks. We take this approach so that we can incorporate high-resolution temporal data and identify high-resolution temporal relationships between events given our access to high resolution Learning Management System (LMS) data, and inspired by the remark by \cite{Mahzoon2018} that \textit{`Access to LMS data makes finer-grained node segmentation possible, which may lead to more timely assessments of academic risk'}. Here we show that high resolution (task-level) data can be informative of student performance.  
%We also point to previous analyses that have highlighted the need for high resolution temporal data and appropriate models\citep{Ye2014,Ye2015,Tang2016,Mendez2014,Liu2018}.

%Therefore, in this paper we introduce new methods of analysis and modelling for temporal sequence data. We then show how these data-driven and model based analyses can provide insights into learning data and can identify anomalous patterns for further examination or predict performance. 

%Indeed, \cite{Ye2014,Ye2015} show that finer-grained temporal information improved predictive power of student performance. 

% Additionally, the temporal sequence data framework is flexible; we are able to analyse student trajectories and also analyse course learning design.
% %Moreover, whilst our model can be used to analyse student success, we also show that our model provides an avenue for identifying anomalies in learning design.
% %Beyond prediction of drop out or performance, temporal analyses are important for guiding learning design and re-design.
% A study by \cite{Mendez2014} concluded that using analytical techniques, including temporal analyses such as sequence mining, could pinpoint interesting anomalies or irregularities for a more focussed investigation by individuals with deep knowledge of the course. Accordingly, we can use our model to take task centric perspective and identify common irregularities in task sequences across students.

In this paper, we leverage LMS data and a sequence data framework to introduce both novel data-driven analyses and a Bayesian probabilistic model for the purpose of analysing and modelling sequence data. The data-driven analyses provide insights into our learner engagement data; for example, we show that the temporal data resolution (task level vs. weekly coarse grained sequences) can lead to complementary but distinct conclusions about learner engagement. The Bayesian probabilistic model is used to model student trajectories and to predict student performance as a function of sequence length, i.e., how early can we predict student performance from their trajectory.

The paper is structured as follows: In Section \ref{sec:methods} we discuss the mathematical notation and structure of the data sequence framework. We also introduce methods for analysing the data sequences of students and the Bayesian probabilistic model for modelling sequence trajectories. In Section \ref{sec:results} we explore the ensemble data sequence trajectories of a student cohort undertaking an on-line course. In Section \ref{sec:taskcompletion} we consider the patterns and engagement behaviours of the ensemble. We then compare the sequence engagement behaviours of students at high-resolution (task level) and coarse-grained (weekly) and compare the engagement behaviours of high performance and low performance groups. In Section \ref{sec:hypertraps} we apply the Bayesian probabilistic model to high and low performance groups to learn sequence trajectories associated with performance. Finally, in Section \ref{sec:taskcentric} we take a task-centric perspective and look at how the sequence data model can be used to reveal quantitative information about individual tasks and their relationship with aggregate student performance.

In summary, our paper provides the following contributions:

\begin{itemize}
 \item Within the framework of sequence analyses in learning analytics,  we construct a data sequence method that treats single tasks as nodes, thus allowing the exploration of temporal relationships between tasks. 
 \item We compute key quantities that describe the probability of learners transitioning between tasks, and use these metrics to identify anomalous learner behaviours.
 \item Comparing high-resolution temporal data with coarse-grained temporal data, we show that the temporal resolution of the data can lead to distinct conclusions about student engagement behaviour and its relationship with performance.
 \item We introduce a novel Bayesian probabilistic model that is capable of modelling trajectories, and combine this model with a Bayes' classifier to predict student performance.
 \item In contrast to student-focused temporal methods that extract temporal features about student completion to predict student success, we show how the data sequence framework can also introduce metrics to analyse course design.
\end{itemize}

\section{Methods}\label{sec:methods}

In this section we first describe and summarise the high-resolution dataset used to exemplify our analytical methods. Secondly, we describe the structure of the sequence data framework which forms the basis of the analytical methods and models introduced in this paper for the analysis of temporal sequence data. We then highlight key data-driven quantities that can be extracted from our data sequence model representation for the analysis of high-resolution LMS data. We then introduce a Bayesian probabilistic model for modeling sequence trajectories of learners and use it for predicting performance. Finally, we show that sequence data framework can be used to perform task-centric analyses and introduce data-driven metrics to reveal insights into course learning design.

%We describe how the high-resolution data can be coarse-grained into node that represent weeks instead of individual tasks, a necessary step to highlight the importance of high-resolution data. Finally, we introduce a `confidence' metric that allows us analyse relate task-level confidence data such that we can attempt to understand anomalous task completion sequences.

\subsection{Course information}

\begin{figure}
\begin{floatrow}
\ffigbox{%
  \includegraphics[width=0.4\textwidth]{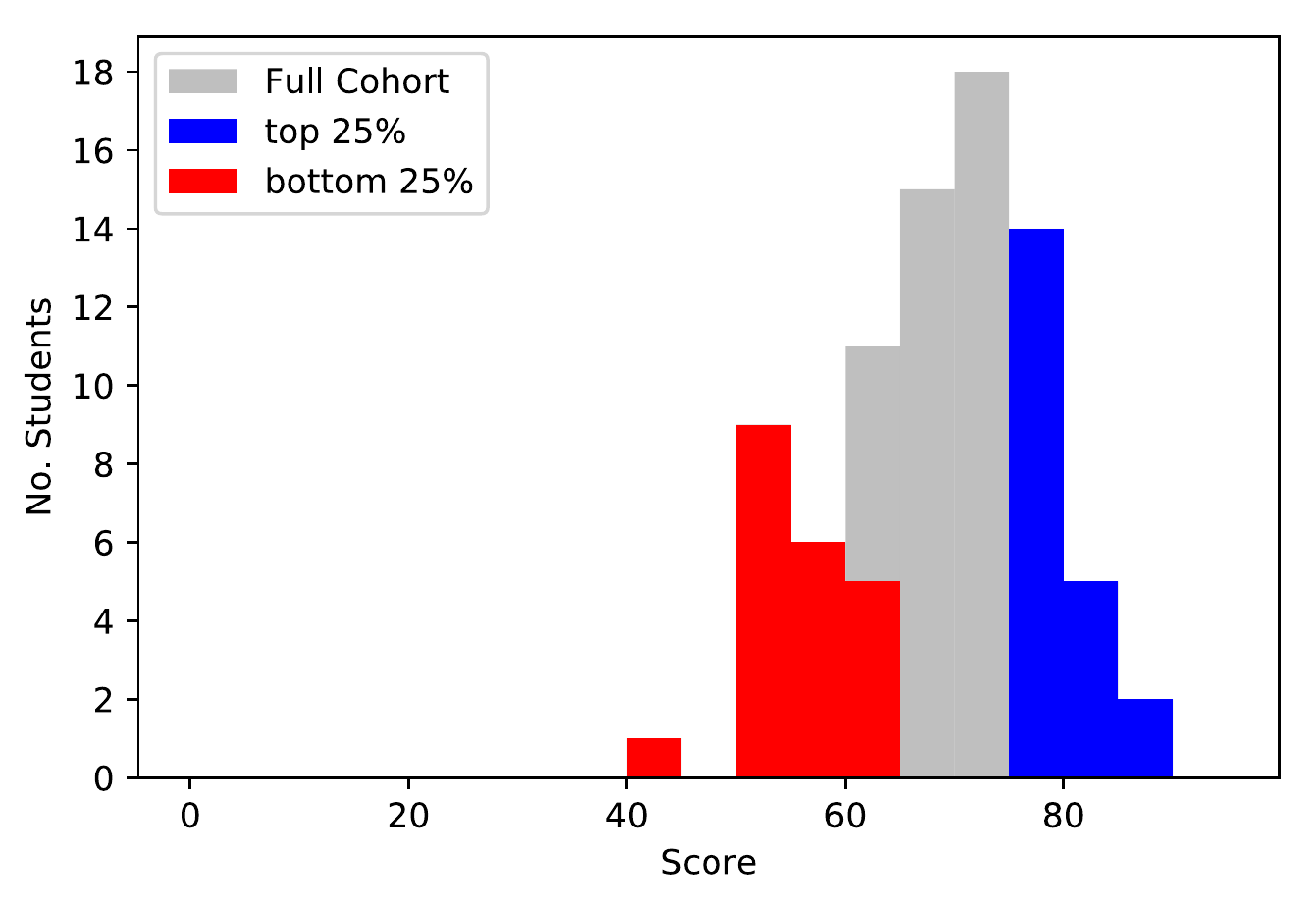}%
}{%
  \caption{The grade distribution for the entire cohort with the bottom and top 25\% groups highlighted in red and blue respectively.}%
  \label{fig:performancedist}

}
\capbtabbox{%
\begin{tabular}{ll}
\hline
Feature            & Value               \\ \hline
Population size, N & 81                  \\
Module name        & Corporate Finance   \\
Degree Course      & Online MBA (2 year) \\
Male/Female ratio  & 57/24               \\
No. Tasks, T       & 123                 \\
No. Sessions       & 10                  \\
Mean Grade  (.std)     &   68.4\% (9.3\%)                \\
Mean Grade bottom 25\% (.std)       &   55.6\%   (4.6\%)             \\
Mean Grade top 25\% (.std)        &     79.3\%  (3.3\%)           \\

Age range          & 28-53               \\ \hline
\end{tabular}
}{%
\caption{Table of summary statistics for the dataset used in this paper.}
\label{tbl:summary}
}
\end{floatrow}
\end{figure}

The subjects in this study were $N=81$ post-experience learners pursuing a 2-year post-graduate part-time Management degree at a selective research orientated institution, a summary of the data is displayed in Table \ref{tbl:summary}. 
%The learners formed part of a total cohort of 87 learners; the 
%data from the remaining 6 learners was not included as these learners either interrupted their studies or withdrew from the programme. 
The cohort ranged in age from 28 to 53 years old (57 males, 24 females) and resided in 18 geographically disparate countries.
Although the subjects met face-to-face at the start of each academic year, the course was studied completely online without any further physical interaction.
We gathered computer interaction data of the learners undertaking the online management course `Corporate Finance' during the second semester of the first academic year. The anticipated study load was 5 to 7 hours per week for the single course we analyzed. This course was run in parallel with a second course (not analyzed here) with a similar workload. An online session was delivered to the students each week, with the course running over 10 weeks (for a total of 10 sessions). The course was assessed via a combination of coursework and a final exam. Coursework was undertaken at three points during the course: Sessions 3, 4 and 9. The performance distribution of the entire cohort had tail to low performance with a median of 69.7 (Figure \ref{fig:performancedist}). 
% \begin{table}[]
% \centering
% \begin{tabular}{ll}
% \hline
% Feature            & Value               \\ \hline
% Population size, N & 81                  \\
% Module name        & Corporate Finance   \\
% Degree Course      & Online MBA (2 year) \\
% Male/Female ratio  & 57/24               \\
% No. Tasks, T       & 123                 \\
% No. Sessions       & 10                  \\
% Age range          & 28-53               \\ \hline
% \end{tabular}
% \label{tbl:summary}
% \caption{Table of summary statistics for the dataset used in this paper.}
% \end{table}

\begin{figure*}[!t]
    \centering
    \includegraphics[width=0.8\textwidth]{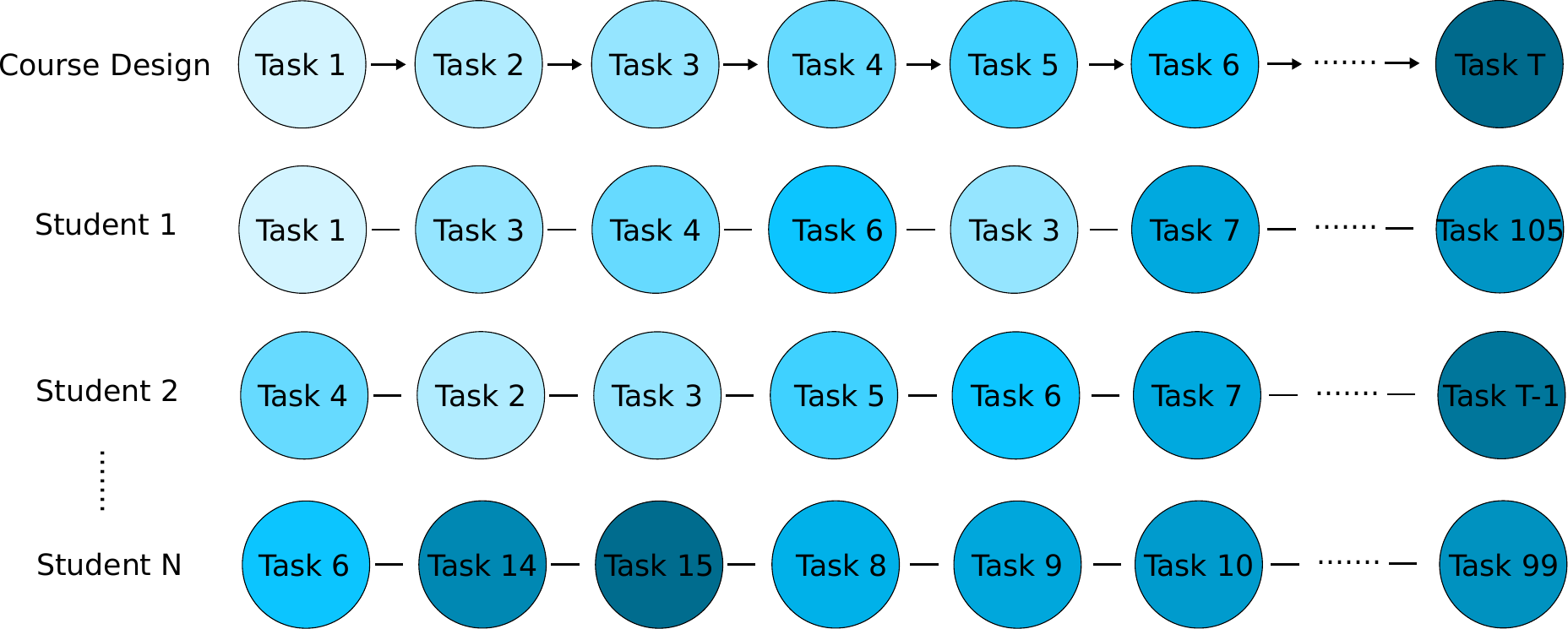}
      \caption{The sequence data framework represents each task as a node which is ordered according to the course learning design. Each student is represented by a sequence of tasks. This framework forms the basis for application of various data-driven analyses and Bayesian sequence modeling.
      }
    \label{fig:sequenceframework}
\end{figure*}

\subsection{Sequence data framework}

The 10 sessions of the course are made up of a total of $T=123$ tasks identified by their 'Task ID', an integer from 1 to 123 representing the \emph{nominal order} of the course layout, i.e., the order in which tasks appear to learners as they navigate the online course website.  
This nominal task order $g$ is represented as the ordered set $s^{(g)} = \{1,2,...,123\}$.
Each learner $k$ is described by the ordered set of completed tasks: %$s^{(k)}$ 
%At the beginning of the course, each learner starts with no completed tasks and their set of  $s^{(k)}=\{\}$. As they complete tasks, each learner
%The actions of each learner $k$ are represented as an ordered set of completed tasks:
$$
s^{(k)} = \{t^{(k)}_1,...,t^{(k)}_n\}
$$
%The state of the $k^{th}$ learner before having completed any tasks may be represented as the empty set $s^{(k)}_{i=0}=\{\}$. 
% $$
% s^{(k)}_{i=n} = \{t^{(k)}_1,...,t^{(k)}_n\}
% $$
where $t^{(k)}_j$ is the task ID of the $j^{th}$ task completed by learner $k$. 
%At the beginning of the course, the $k^{th}$ learner before having completed any tasks is represented as the empty set $s^{(k)}=\{\}$. 
%The complete ordered set of tasks for the $k$-th learner is referred to as $s^{(k)}$. 
Note that the sets may be of different lengths for different learners since task completion is not compulsory and the number of completed tasks may vary from individual to individual. Figure~\ref{fig:sequenceframework} displays some example sequences for students using the framework described here.

We can also coarse grain our sequence framework. Given that tasks are grouped into sessions we compute the transition probability  $p(\text{next session } s+1|\text{ previous session }s)$, by substituting the task IDs for session IDs in the ensemble $S$. We use all these quantities to analyze the behaviors and critical juncture tasks associated with the course.

\subsection{Data-driven learner centric analysis}

Given our data sequence framework we introduce two key methods for analyzing the sequence data to investigate learner behavious. The whole cohort of learners is then described by the ensemble of ordered sets of tasks, $S = \{s^{(k)}\}_{k=1}^{81}$. From this ensemble, we can compute the following two key quantities:
\begin{enumerate}
    % \item  \textcolor{orange}{The probability $p_{ij}$ that task $i$ is $j^{th}$ element of learner sequence $s^{(k)}$ } \MB{I think this is not right... it is not for a particular $k$?? I think it is just averaged over all $k$?? This is my proposal:}.
    
    \item The number of times that task $i$ appears as the $j^{th}$ element across the ensemble $S$ is denoted as $n_{ij}$. The probability $p_{ij}$ that task $i$ appears as the $j^{th}$ completed task across $S$ is thus given by $p_{ij} = n_{ij}/\sum_j n_{ij}$. These probabilities are compiled in the $T \times T$ matrix $P=(p_{ij})$.
    
    \item To quantify the transitions between tasks, we count the number of times $m_{ij}$ that task $j$ is immediately preceded by task $i$ across the ensemble $S$, and normalise it across all transitions to obtain the probability that task $j$ is preceded by task $i$ across the ensemble: $p_{i \to j} = m_{ij}/\sum_{i,j}m_{ij}$. 
    To obtain the conditional probability, we need to normalize across all tasks:  $\pi_{ij}= p(\text{next task} j |\text{ previous task } i) = p_{i \to j} / \sum_j p_{i \to j}$. These conditional probabilities are compiled in the $T \times T$ matrix $\pi= (\pi_{ij})$.
    %(p(\text{next task} j |\text{ previous task } i))$. 
    It should be noted that this matrix provides a summary of transitions to any task %$j$ 
    given a previous task acquisition, %$i$, 
    but $\pi$ is not a stochastic transition matrix for task acquisition due to tasks being irreversibly acquired.
    
    % $\pi_{ji}= p(\text{next task} j |\text{ previous task } i)$.  \MB{I changed this also. Check that it is precise. Is it $\pi_{ij}$ or $\pi_{ji}$ in the definition above??}. These probabilities are compiled in the $T \times T$ matrix $\pi=(\pi_{ij})$.
\end{enumerate}
%, with this matrix summarized as a histogram. 
%Secondly, we compute the probability $p(\text{next action}|\text{ previous action })$ 
%that a given task $j+1$ is immediately preceded by task $j$. 
%can also be computed, which we refer to the corresponding probability as $p(\text{next task }j|\text{ previous task }i)$. 
%The `start' and `end' points are tagged to facilitate the description of the first and last tasks.
%and the task that preceded completing the course. However, we do not explicitly include them within the set. 
%\MB{So is it true that $s^{(k)}_{i=0}=\{\}$ or is it $s^{(k)}_{i=0}=\{ 'root' \}$ ?? If not, we should rewrite slightly...}

The sequence analysis methods were adapted from \cite{Greenbury2018}.

\subsection{Bayesian probabilistic model: a generative model for task sequences}
In the description of a learner's task sequence, no assumptions about an underlying model that may have generated the sequences. The most general probabilistic model assumes that task completion is determined by all the tasks previously completed. If we assume that the order in which previous tasks were completed can be ignored, this model is a first-order Markov chain with a hypercubic state space defined by $2^T$ states and an associated $T \, 2^{T-1}$ edges representing the transition matrix between these states. As $T=123$, the state space and associated parameter space of the transition matrix is too large to be tractable.

Here we introduce the application of  HyperTraPS, a recent probabilistic Bayesian model described in \cite{Greenbury2018,johnston2016evolutionary}. 
We fit this model to derive parameterisations  that can be used for student classification. The model reduces the full hypercubic state space to a tractable $T^2=15,129$ state space by assuming that the probability of the next task in the sequence is proportional to a basal rate of acquisition for that task and a independent contributions from tasks already completed. These are fitted from the aggregate set of observed transitions (the student task sequences).

As we have complete information on the order in which tasks are completed, the utility of a generative model in this case may be found in its use as a classifier. %rather than for learning the order in which tasks are completed in the absence of all task completion information. 
Utilising and extending the Na\"ive Bayes classifier introduced in~\citep{Johnston2019}, we consider the probability that a learner $k$ belongs to the high performing group $g_1$ and the low performing group $g_2$ after completing $n$ tasks with task set $s^{k}$. We can write the odds ratio relating to this probability as:
$$
\frac{P(k \in g_1 | s^{(k)})}{P(k \in g_2 | s^{(k)})} = \frac{P(s^{(k)} | \pi(g_1)) P(g_1)}{P(s^{(k)} | \pi(g_2)) P(g_2)}
$$
where $\pi(g_j)$ are posterior samples drawn from the HyperTraPS model fitted to a set of sequences from labeled group $j$. 
In this paper, the two groups considered are the top 25\% ($g_1$) and the bottom 25\% ($g_2$) based upon their score in the course.

\subsection{Data-driven task centric analysis}

Given the sequence data framework we are able to computer compute statistical properties of every task: the frequency with which it is completed; the mean rank across all learner sequences; and the difference between $p(\text{next task }|\text{ previous task })$ across different sub-groups.

\subsection{Confidence analysis}

To investigate the effectiveness of our analytical methods and model, we use a student confidence analysis to introduce a `story-telling' or causal inference aspect . At the end of each session, students are asked to reflect on the tasks they had been given to complete. There were given three options to choose from for each task: (a) 'Yes I feel confident I can do this', (b) 'I need to revisit this', and (c) 'I need more support'. For each task we calculate the proportion of students that felt confident (stated confidence) or did not feel confident (needed to revisit or requested support).

Using this data we identify when a student was not confident on an individual task, and we measure the confidence of a student across a collection of tasks,
$$
\mathcal{C} =\frac{\textit{`confident'}}{\textit{`revisit' + `support' + `confident'}}
$$
where a larger value indicates a more confident student and a lower value indicated a less confident student. Using this additional data, we can associate anomalous sequence orderings with measures of student confidence at both individual and group levels.

%------------------------------------------------

\section{Results and Discussion}\label{sec:results}

\subsection{Data-driven learner centric analysis}

\paragraph{Ensemble task completion behaviors of student cohort relative to course design}\label{sec:taskcompletion}

\begin{figure*}[!h]
    \centering
    \includegraphics[width=0.8\textwidth]{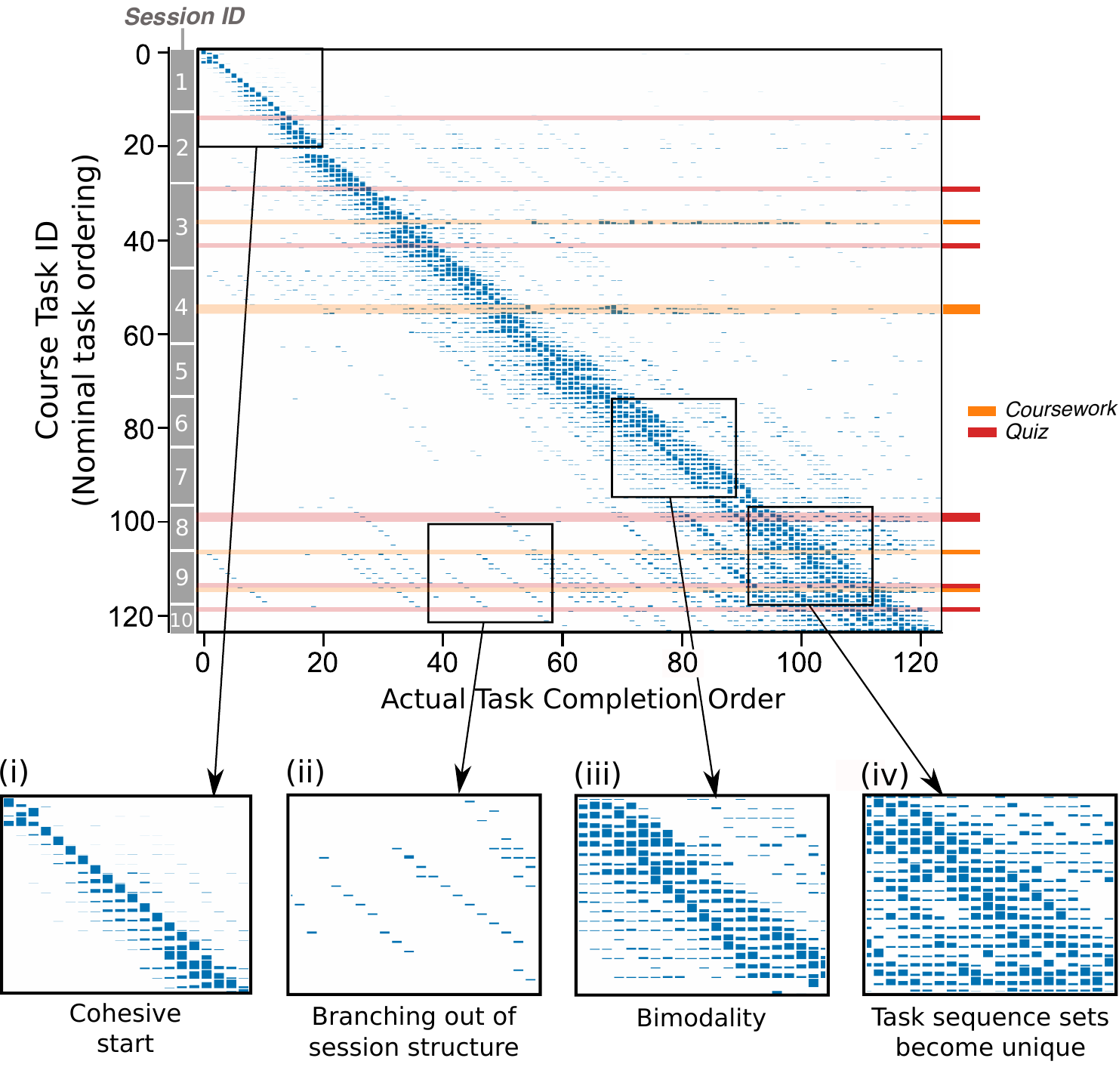}
      \caption{(\textit{A}) 2D-histogram compiling the probabilities of task completion for the 81 learners. The $y$-axis denotes the course task ID, i.e., the nominal completion order. Each task is part of a weekly session. The $x$-axis displays the point in the learners' sequence in which that task was actually completed. Some of the key task types (coursework, quizz) are indicated by colours. (\textit{B}) Each insert shows a magnification of a different part of A: (i) the beginning of the course;  (ii) large deviations from the nominal task order later on; (iii) bimodality of responses for particular tasks; and (iv) the broad distributions of task completion towards the end. 
      }
    \label{fig:1a}
\end{figure*}

As defined above, the completion sequences of $N$ learners over $T$ tasks can be summarized as a $T \times T$ 2D-histogram (or matrix of numbers) $P$, where each element $p_{ij}$ corresponds to the probability that task $i$ was completed in the $j^{th}$ position across the ordered sequences of the cohort. Figure~\ref{fig:1a}A displays this histogram for our learner cohort, with tasks ordered according to their nominal task order $g$ ($1-123$) along the vertical axis and plotted against the actual completion order in the learners sequences on the horizontal axis. 
The histogram for each task across each row summarizes the spread of the actual order of completion across the learners in the cohort. 
If all learners had completed \emph{all tasks in the nominal order $g$}, then $P$ would have been the identity matrix, with all the probability located on the diagonal. Therefore, the off-diagonal spread signals departure from the nominal order across the cohort. 
For example, learners completed task 1 as their first task with high coincidence (77\%, $p_{11}=0.77$), yet, even for this first task, there was a sizeable proportion of learners (21\%, $p_{12}=0.21$) that completed task 1 as their second task.
Overall, the strong diagonal component of Fig.~\ref{fig:1a}A suggests that learners generally follow the nominal task order, with deviations evidenced by the presence of such off-diagonal elements.

At the beginning of the course, we observe a cohesive start in which
learners generally follow the expected course structure as evidenced by the sequential peaks of each histogram up to task 15 (Fig.~\ref{fig:1a}B(i)).
Yet, more in detail, we observe some deviation from the nominal course structure. For example, very early on task 2 is most commonly completed in third position, whilst task 3 is completed second (Fig.~\ref{fig:1a}B(i)). Considering these two tasks in more detail, task 2 requires the learner to complete a vast amount of reading whilst task 3 requires the learner to submit an estimate for a quantitative question. Task 3 is located on the next screen requiring the learner to actively leave the web page containing task 2 to access task 3. Since task 3 is inherently related to task 2 (both explore financial inter-mediation), it could be that learners did not feel they comfortably understood the material in task 2 until they completed task 3. %Interpretation and reasoning of particular tasks and their ordering is not the purpose of this research.
%After the deviation from nominal task order associated with task 2 and task 3, we observe that learners generally follow the expected course structure as evidenced by the sequential peaks of each histogram from task 4 to task 15 (see Fig.~\ref{fig:1a}B(i)). 
Another deviation occurs between task 15 and task 16 
(Fig.~\ref{fig:1a}B(i)), whereby task 16 is completed earlier than task 15 on average. Task 15 requires the learner to complete a quantitative quiz with several complex financial questions. Although the quiz did not contribute to their final course grade, the learners may have wanted to leave the quiz until the end of the learning session, or they might have believed that additional material may become available later to aid them in the quiz. 
%However, the quiz did not contribute to their final course grade.

%We are also able to observe this coherent start in Fig.~\ref{fig:1b}B, whereby the majority of learners complete tasks that are similar to the nominal task order. As expected, the majority of learners begin by completing tasks at the beginning of the course.  

Beyond such small deviations from the nominal task order, we also identify much larger deviations. Fig.~\ref{fig:1a}B(ii) highlights tasks that have been completed much earlier that expected from the nominal task order. Interestingly, these tasks are completed in sequential order within their session, signalling a jump-ahead from the learners to an out-of-order later session.
Such jump-ahead deviations appear as diagonal streaks in the lower triangle of Fig.~\ref{fig:1a}A, but note that similar deviations are also observed in the upper triangle of Fig.~\ref{fig:1a}A indicating tasks that were completed later relative to the nominal task order.

There are several features of the cohort dynamics that become more prominent as task completion progresses. In Fig.~\ref{fig:1a}B(iii), we observe an example of bimodality (two peaks) in task completion indicating that there are two groups of learners completing these tasks systematically, but at different points in their sequences (relative to the nominal task order). Note that the presence of bimodality is not simply an artifact of learners randomly missing tasks, if so we would observe a single broad peak, but the emergence of subgroups of learners following different strategies. Indeed, this instance of bimodality appears at the beginning of Session 7 (task 84--task 95) and becomes less pronounced after Session 7 is finished.  
After a more thorough analysis, we find that a large group of learners appear to skip both task 78 and task 81 whilst the remaining learners complete these two tasks, resulting in a branching of two sets of learners corresponding to left-hand and right-hand peaks. Both of task 78 and 81 are interactive tasks that required an application of learned knowledge, which may have caused the split between those that were able to complete the tasks and those that were not. 

As task completion progresses, we observe a broadening of the task distributions (Fig.~\ref{fig:1a}B(iv)).
Such broadening is expected when completion of tasks is not mandatory; if a learner misses a task and continues to follow the nominal task ordering of the course they will have a shift in their actual completion sequence relative to other learners.
%i.e. following the nominal task ordering but skipping task 3 will give the sequence $s^{(j)} = \{t_1=1,t_2=2,t_3=4,t_4=5,t_5=6,...,t_{T-1}=T\}$. In fact, missing any number of tasks will cause a systematic shift in the distribution. 
The broadening is amplified towards the end of the course, as a consequence of learners skipping an increased number of tasks (variable across the cohort of learners), in order to finish the course before the final exam. 

\begin{figure*}[!h]
    \centering
    \includegraphics[width=0.7\textwidth]{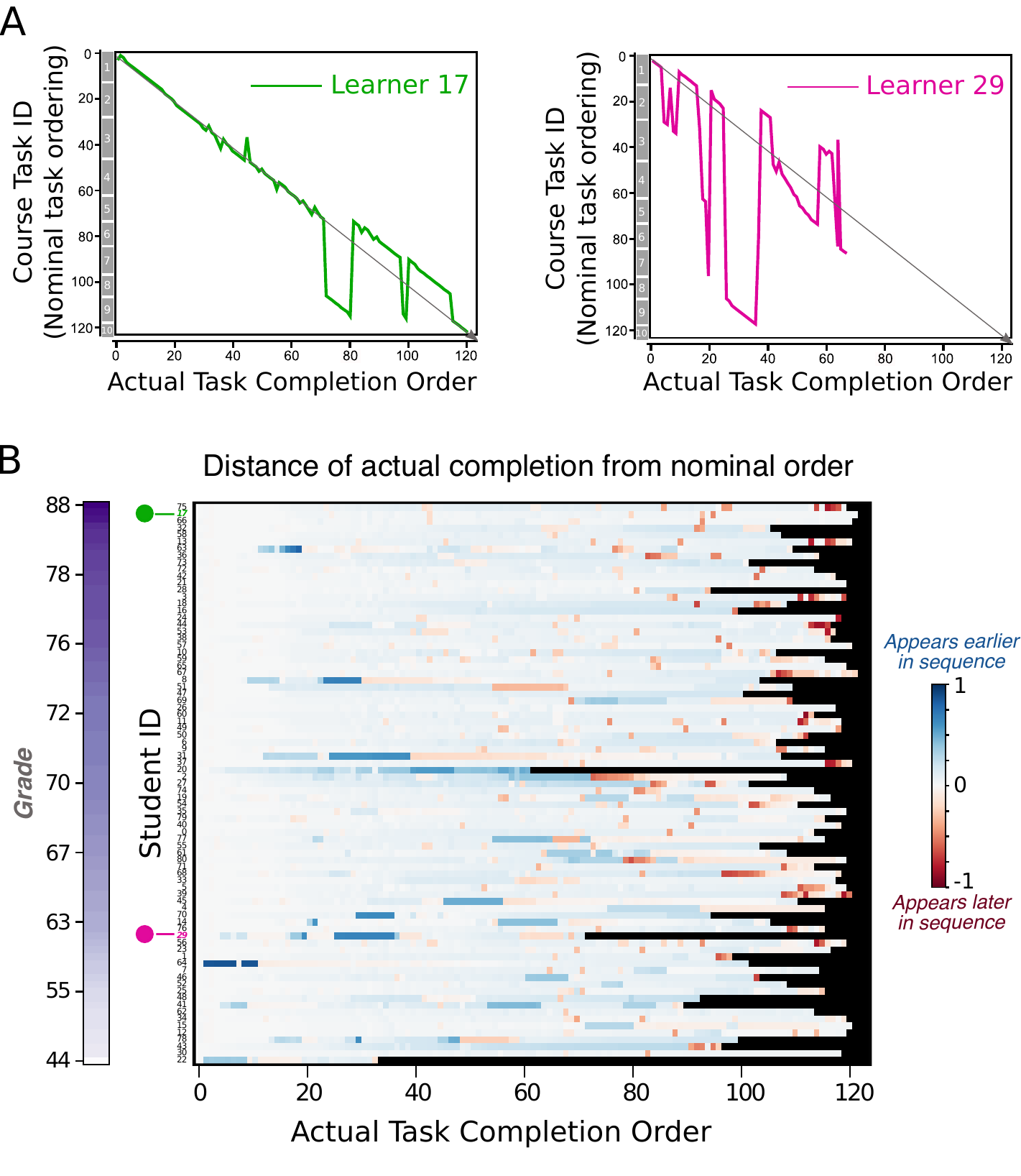}
      \caption{(\textit{A}) Sample trajectories of two learners (17 and 29) as they undertook the course. Learner 17 followed closely the nominal task ordering for the majority of the course only deviating towards the latter tasks. Learner 29 exhibited large deviations from the nominal task ordering throughout and only completed half the course. (\textit{B}) A heat map displaying the relative distance between the actual completion order and the nominal order for all learners (rows). The learners are ordered by descending performance from top to bottom, as indicated by the grade colourmap. If a learner had completed the course exactly as the nominal order dictates then the row would be white. Tasks completed earlier relative to the nominal order appear blue, whilst tasks completed later than the nominal order appear red. The region after the student has completed the course is colored in black. The learners in A are indicated by a green dot (learner 17) and a purple dot (learner 29).
      }
    \label{fig:1b}
\end{figure*}

An examination of the deviation of the different learners from the nominal task order is presented in Fig.~\ref{fig:1b}. 
Fig.~\ref{fig:1b}A shows two distinct task sequences for two learners: learner 17, who follows closely the nominal course order almost all the way to completion, and learner 29, who completes only about half of the total tasks in a more idiosyncratic order, with several jumps between sessions and completing only a little over 50\% of the tasks of the course. 
%Two task completion trajectories of learners 17 and 29 are displayed in Fig.~\ref{fig:1b}A. We observe that learner 17 generally follows the nominal task ordering for the early stages of the course whilst learner 29 exhibits large deviations from the nominal task order from the beginning of the course. 
%Learner 29 first completes about 50\% of session 1 before jumping directly to session 3 to complete a few tasks and then proceeds to return to session 1. These large deviations continue throughout the course; we observe learner 29 make transitions from session 2 to session 8, session 9 to session 2, and session 5 to session 3. Moreover, learner 29 only completes a little over 50\% of the course.  
We compile this information for all 81 learners in Fig.~\ref{fig:1b}B, where we show the (normalized) distance between the actual completion of tasks and the nominal task order for each learner ordered in descending order of performance from top to bottom. Tasks in blue were completed earlier in the sequence relative to the nominal order, whilst tasks in red indicates later completion in the sequence relative to the nominal order. 
The learners of Fig.~\ref{fig:1b}A are indicated by dots (green for learner 17, purple for learner 29). The observed deviations of learner 29 are seen as blocks of tasks (in blue and red) completed out of sync with the nominal task order. In contrast, the profile of learner 17 is almost totally white, indicating a constant progression in accordance with the nominal order. 
Fig.~\ref{fig:1b}B also shows that low performing learners tend to exhibit large deviations from the nominal task order more often than high performing learners (although certainly not exclusively) and a lower fraction of completed tasks (as evidenced by the larger black blocks). 
Note that the figure shows sequential blocks of tasks, commonly corresponding to a single session, which have been completed out of sequence relative to the nominal order.
This indicates that analyzing sequences not only at the level of tasks, but also at the level of weekly sessions can provide a meaningful temporal basis for the analysis of learners' behaviors~\citep{Lund2017}, as we explore below.

In summary, the results in this section have shown that using our data sequence model we are able to analyse the engagement behaviors of students at a high-resolution level. We are able to identify both anomalous deviations from the designed course structure at both a cohort level and for individual students. We have shown that once we have identified anomalous patterns, such as that for learner 29, we are able to provide a more indepth analysis that may reveal causal reasons.

\begin{figure*}[!t]
    \centering
    \includegraphics[width=0.75\textwidth]{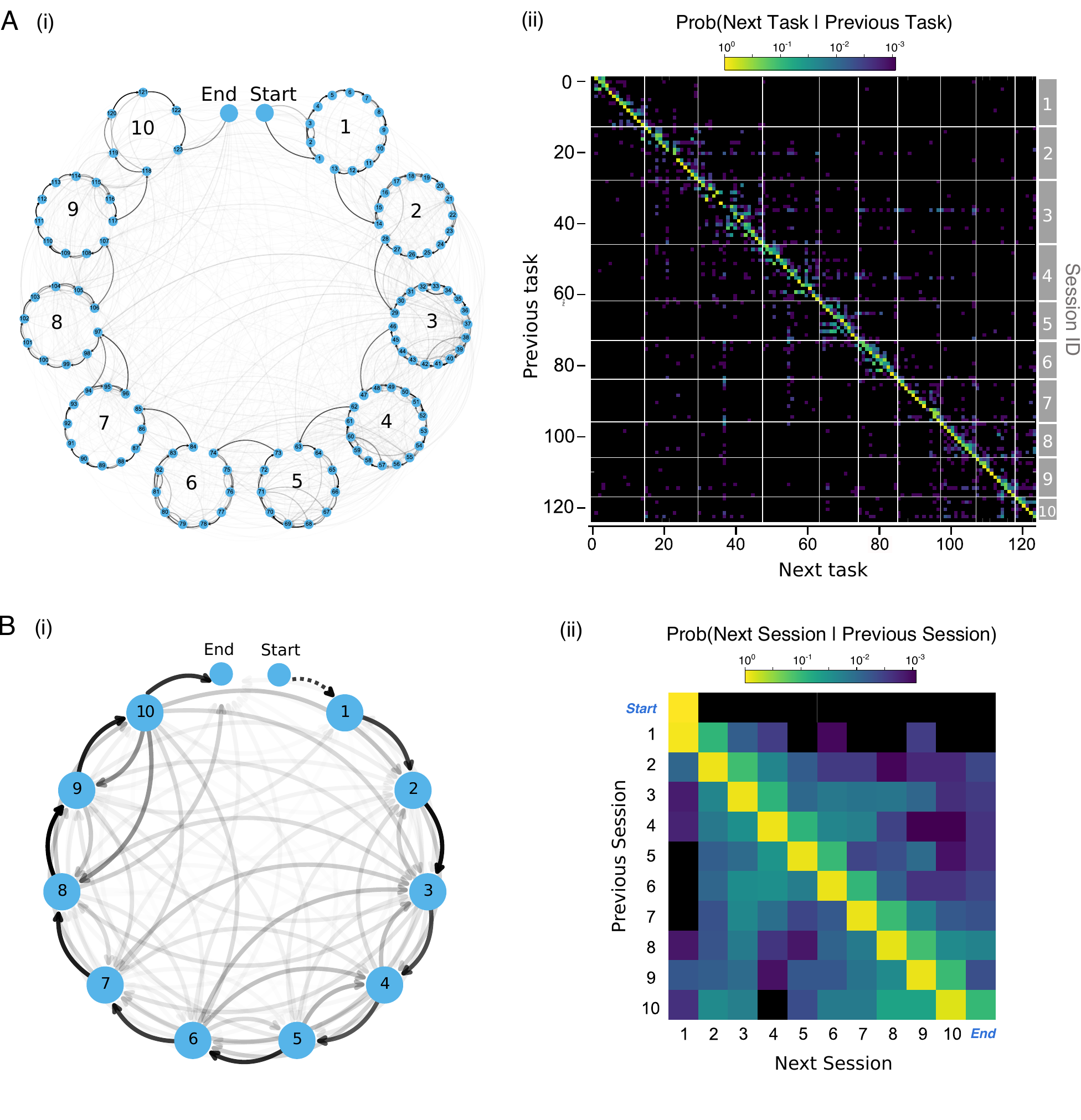}
    \caption{(\textit{A}(i)) The transition probability between tasks $\pi$ is represented as a transition graph. Each blue node corresponds to a task, each sub-circle corresponds to a weekly session of tasks (numbered from 1-10), and the entire circle corresponds to the full course. The edge thickness between tasks corresponds to the probability $\pi_{ij}$ 
    %\MB{$\pi_{ij}$ or $\pi_{ji}$???} \SG{Yes $\pi_{ij}$ is right as first index denotes rows (previous task) and second index is columns (next task)} 
    of transitioning to task $j$ given the previously acquired task $i$. Generally, there is a high probability of transitioning between sequential tasks; however, there are often large non-sequential transition probabilities. (\textit{A}(ii)) A heatmap of the matrix $\pi$ (in logarithmic scale). Off-diagonal elements correspond to deviations from the course structure.  (\textit{B}(i)) A transition probability graph between weekly sessions, where each node corresponds to a session and the edges between sessions $i$ and $j$ are weighted by the probability %$p(j|i)$ 
    of transitioning between any task in session $i$ to any task in session $j$. 
    %(The transition probability from start to session 1 is a magnitude larger than other transitions and is clipped for visualisation purposes). 
    Although there is a strong probability of sequentially completing the sessions, there are is a significant number of deviations from the session sequence (e.g., from session 10 to sessions 9 and 8, or from session 6 to session 3). The transitions within a session (self-loops) and the transition from start to session 1 are not shown for clarity of visualisation (\textit{B}(ii)) A heatmap of the transition probability between sessions. 
    %where each element corresponds to the probability of transitioning from session $i$ to session $j$. 
    The diagonal exhibits the highest probability given that a learner is most likely to transition between tasks within a session followed by the upper diagonal, representing transitions to the next session in the nominal sequence.
    }

    \label{fig:2}
\end{figure*}

\paragraph{A comparison between high-resolution and coarse-grained temporal sequences}\label{sec:highres_ensemble}

Above, we extracted quantities that revealed deviations from the designed course sequence for both individual students and the ensemble of students. In this section we explore the high resolution task-level data and compare it against a coarse grained version of the data (coarse grained by week).

Fig.~\ref{fig:2}A plots the task transition probability matrix $\pi$ defined above in two different (but equivalent) ways: as a transition graph (i) and as a transition matrix (ii).
%To summarize the sequence information, we plotted the task transition probability network for all learners in Fig.~\ref{fig:2}A(i) and the corresponding transition matrix in Fig.~\ref{fig:2}A(ii). 
%In Fig.~\ref{fig:2}A(i) each task is grouped into its respective weekly session of which there are 10 in total. 
As expected, we find strong transitions that follow the nominal structure both within each session as well as clockwise from session to session in Fig.~\ref{fig:2}A(i). 
This is similarly shown in Fig.~\ref{fig:2}A(ii) by the strong concentration on the diagonal and upper diagonal, which indicates the high probability that learners follow the nominal order of the course.  However, we also find a large number of non-sequential transitions between tasks as a consequence of learners jumping forwards or backwards in the course. These are shown as non-sequential edges in Fig.~\ref{fig:2}A(i) and off-diagonal elements in Fig.~\ref{fig:2}A(ii). 
Specifically, we observe a large number of non-sequential transitions within sessions 2, 3 and 5. 
For example, within session 3 there is a large probability that learners complete task 42 and then return to task 38. In this particular case, task 38 is a discussion post where learners were required to publicly write a response to a tutor question whereas task 42 is a quiz with various questions regarding equity markets and valuations. It appears that a large number of learners completed the quiz before the discussion post, perhaps because they did not feel confident completing the public discussion post until they had gained additional information or understanding from the quiz. To explore this further we analysed the  percentage of students that felt confident undertaking this task. We found that only 60\% (59/81) of students were confident in this task (self-assessed at the end of each session), whereas across the remaining tasks within session 3 the average number of students that were confident was higher (69\%, std. 7\%) and the average task confidence across the module was (68\% std. 8\%).   %In Fig.~\ref{fig:2}A(ii) we can also see various off-diagonal elements that indicate non-sequential task completion relative to the nominal task order.

\begin{figure*}[!t]
    \centering
    \includegraphics[width=.75\textwidth]{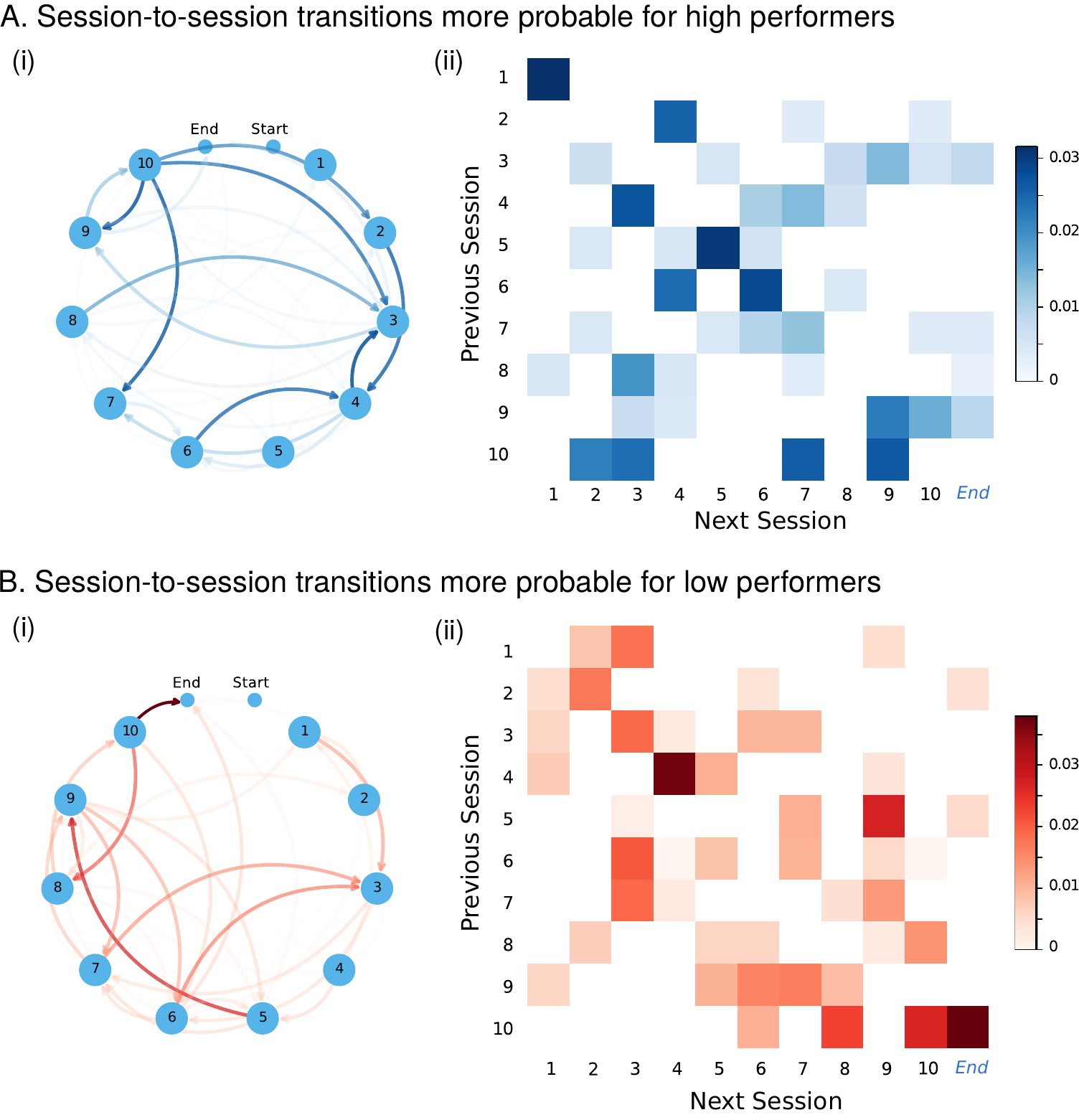}
    \caption{The transition probability between weekly sessions is calculated separately for the top 50\% performers and the bottom 50\% performers and the difference between the probabilities in the two groups is calculated. (\textit{A}) The session-to-session transitions that are more probable for high performance learners (i) mapped onto a network structure and (ii) visualized as a transition matrix heatmap.
    (\textit{B}) The session-to-session transitions that are more probable for low performance learners (i) mapped onto a network structure and (ii) visualised as a transition matrix heatmap.
%\begin{comment}
%top = 0.06150    0.90867    0.02983
%bottom = 0.05364    0.91759    0.02877
%(top-bottom)./((top+bottom)/2) = 13.6%   -1%    3.62%
%\end{comment}
}    
    \label{fig:3}
\end{figure*}

For comparison, we take a coarse-grained approach and calculate the transition probabilities between the 10 weekly sessions (Fig.~\ref{fig:2}B)
represented as: 
%There are a total of 10 sessions that the 123 tasks are divided among. Earlier we found that these sessions represented meaningful analytical units of time for differentiating learner behaviors. These sessions begin with tasks: 1, 14, 29, 47, 63, 74, 85, 97, 107 and 118. 
the inter-session transition graph in (i) and the coarse-grained transition matrix in (ii).  As in the task level analysis, there is a strong clock-wise sequential component in the graph (i) and a concentration of probability on the main and upper diagonal in the matrix (ii), indicating that learners generally follow the session sequence. However, there are obvious deviations. For example, we observe that there is a high probability that learners will transition back to session 3 from several sessions (6, 7, 8, 9, 10). 
%The transition probabilities are exhibited as a matrix in Fig.~\ref{fig:2}B(ii) where a strong diagonal component is observed which suggests that most task transitions will occur within a session. The second upper right triangle diagonal also has a higher probability suggesting learners commonly transition between sessions in order of their sequence. However, of most interest, are third order effects whereby learners transition between non-sequential sessions. For example, 
Similarly, learners at session 10 (the final session) will often transition to previous sessions (sessions 8 and 9), maybe in an effort to revise and complete parts of the course they had skipped over while completing the course.

These results suggest that task-level (high resolution) and session-level (coarse-grained) sequence data provide true but alternative descriptions of the data that can lead to distinct but complementary conclusions. Within the high resolution analysis we are able to identify individual anomalous tasks (such as task 38), whilst within the coarse-grained analysis we identified an anomalous week (e.g. week 3). This section highlights the importance of using high-resolution temporal data coupled with an appropriate temporal model for analysis.
% we find that learners commonly transition from Session 10 back to Session 8 or 9 (this is also observed in Fig.~\ref{fig:comparison_2groups_2}B(i)).

\paragraph{Comparing sequence completion patterns between high and low performance learners}\label{sec:highres_highlow}

%\newline

Given the differences between the high resolution and coarse grained analyses, we perform a similar comparison whilst exploring student performance. As suggested by Fig.~\ref{fig:1b}B, the pathways and patterns of task completion may differ between high and low performing learners. To explore this hypothesis in more detail, we divide our cohort of learners into two groups: those in the top 25\% (Group 1) and bottom (Group 2) 25\% of performers according to course grade. Although meaningful and more nuanced groups could be created, we opted for a split into the bottom 25\% and top 25\% to ensure a bimodal performance distribution. 
%\MB{Internal note to ourselves: They might ask us to do randomisations/bootstraps by considering a set of random splits into 2 equal groups independent of grade, so as to compare the results against those...}\RP{Agreed - we shall see what they say.}

We compute two separate transition matrices $\pi_{HP}$ and $\pi_{LP}$ from the sequences of learners in Group 1 (high performers) and in Group 2 (low performers), respectively, and obtain the difference between both:
$\Delta \pi = \pi_{HP}-\pi_{LP}$.
%the difference in transition probability between Group 1 and Group 2
%$p_{HP}(\text{next task }j|\text{ previous task }i)$ - $p_{LP}(\text{next task }j|\text{ previous task }i)$ for the high-performance (HP) and low-performance (LP) groups. 
%The transition probabilities are mapped onto the session structure in  
In Fig.~\ref{fig:3}A we show the transition graph (i) and transition matrix (ii) where the high performers have larger probability (i.e., the elements of $\Delta \pi > 0$). Similarly, Fig.~\ref{fig:3}B presents the transition graph (i) and transition matrix where the low performers have larger probability (i.e., the elements $-\Delta \pi > 0$).

We identify a number of differences in the transitions when comparing high and low performers. For example, low performers were more likely to transition directly from session 5 to session 9, which contained a coursework task that needed to be completed, and after completion of session 9 they were likely to transition back to session 6. 
In general, low performing learners have a higher probability of transitioning to a task within the same session (diagonal of Fig.~\ref{fig:3}B(ii)) but also a higher rate of transitions between non-sequential sessions (Fig.~\ref{fig:3}B(i)).
%Figs.~\ref{fig:3}A(ii) \& B(ii) exhibit the same transition probabilities in matrix form. In this representation we can also observe the probability of transitioning within a session (self loops were not displayed on Fig.~\ref{fig:3}A(i) \& B(i)). 
%Low performing learners have a higher probability of transitioning to a task within the same session (this is observed in Fig.~\ref{fig:2}B(ii) along the diagonal). 
%This seems contrary to the previous statement, that low performance learners exhibit a higher probability of transitioning between non-sequential sessions, however, 
These two views are only seemingly contradictory: low performers followed the nominal task order within sessions but did not follow the nominal session order within the course. 
We also observe that the only point where high performers are more likely to depart from the course session structure than low performers is in the jump back transitions form session 10, an indication of an effort to revise and complete missing tasks just before the end of the course. 
%\MB{Check this. It seemed we should say something about high performers too and this feature looks pretty clear in the figure...}\RP{Correct - this was a good spot}
These results reinforce the need for high-resolution data and analysis. Using a coarse-grained approach we would observe low performing students less strictly following the course session structure, however, we would not observe their higher probability of following the nominal task order. 

Following on from the confidence analysis in Section \ref{sec:highres_ensemble}, we found that the low performance group were significantly (paired t-test, p=0.041, alpha=0.05) less confident (no. tasks confident, mean=0.65, std=0.27) than the high performance group (no. tasks confident, mean=0.76, std=0.19) across the entire module. At the group level these confidence values correlate with performance as we would expect. However, these results suggest that less confident students are less likely to follow the nominal session order of the course.

\subsection{Bayesian model for sequence trajectories}\label{sec:hypertraps}

\begin{figure*}[t!]
    \centering
    \includegraphics[width=.9\textwidth]{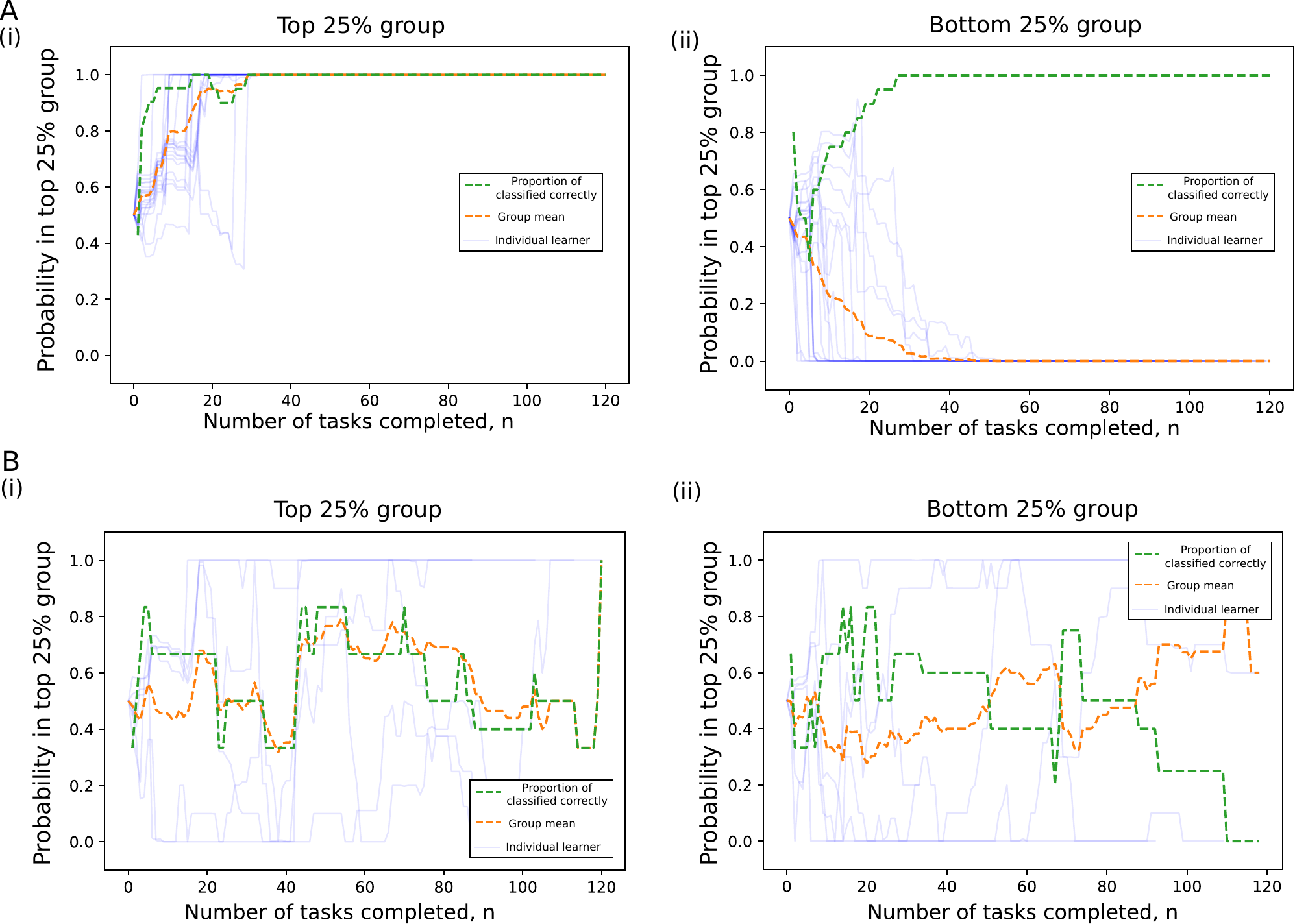}  
	\caption{Application of Bayesian probabilistic model to learn sequence trajectories. (A) Probability that each learner in top 25\% belongs to top 25\% (i) and the probability that each learner in the bottom 25\% belongs to top 25\% (ii) using averages across posterior samples trained on data from the respective groups. Each learner is a blue line, the mean across all learners an orange dashed line and the proportion over 0.5 as a green dashed line. After around 30 tasks the probability goes to unity and zero respectively for the two test groups providing indication that the probabilistic model is able to learn differences between the two groups. (B) The top plots utilized data included in the training dataset for testing, while the bottom plots train with 70\% of the learners from each group and test on the remaining 30\% from each group. This indicates whether the learned model can differentiate unseen learners giving it true predictive utility. There is much greater variability in the position of the each learner's blue line, but on average the orange and green lines lie above and below the 0.5 threshold indicating there is some predictive power. Interestingly, the strongest predictive region seems to be in the gap where coursework is completed later after completing more of the course, a trend observed in the higher performing group. The predictive power weakens at large $n$ due to fewer training and test samples completing that many tasks.}
	\label{fig:hypertraps}
\end{figure*}

In this section, we use the probabilistic generative model to examine whether we can predict the group (top 25\%, $g_1$ or bottom 25\%, $g_2$) to which a student belongs after completing $n$ tasks with task set $s^{(k)}$. We perform two experiments to consider the plausibility of such a model for the data.

In our first numerical experiment, we test whether the probabilistic model can learn parameterisations  that can distinguish two groups. We draw posterior samples $\pi(g_1)$ and $\pi(g_2)$ from HyperTraPS for the entire set of $g_1$ and $g_2$. For each learner in the training datasets $g_1$ and $g_2$, we plot the probability that they belong to their respective groups. Fig.~\ref{fig:hypertraps} (top) shows the result of this when the test learners that we are trying to predict group identity are including in the training dataset. In this case, the left-hand plot (for the top 25\% learners) and right-hand plot (for the bottom 25\% learners) show strong predictive ability after only a few tasks have been completed in the set of all tasks with the probability going to unity and zero for the groups respectively after around 30 tasks. This indicates the probabilistic model is able to learn differences between the two groups in its parameterisations , an important prerequisite for establishing whether learners task set $s^{(k)}$ in conjunction with parameterisations  of a generative model can be used to predict performance.

In our second numerical experiment, we test whether the probabilistic model can use learned parameterisations  to predict which group unseen learners belong to. We do this by splitting our groups $g_1$ and $g_2$ into a 70\% training and 30\% test split. We draw posterior samples from $\pi_\text{train}(g_1)$ and $\pi_\text{train}(g_2)$, and then perform the same method to calculate the probability that learners from the test split belong to $g_1$ and $g_2$ (given we know their true labels). Fig.~\ref{fig:hypertraps} (bottom) illustrates the results for the test split of learners (split by their true label $g_1$ on left-hand side and $g_2$ on right-hand side). The results indicate much less certainty for individual learners, with misclassification occurring. However, the averages across learners in each group indicate some predictive power with average values mainly above 0.5 for $g_1$ and below 0.5 for $g_2$. Additionally, the region of strongest predictive power where the averages are farthest from 0.5 is around the region where we have observed learners being particularly divergent in their choice of task suggesting the results may be robust. Given such small dataset sizes of only around 15 learners for training, this result would be able to be validated on courses with many more learners available.

\subsection{Data-driven learner centric analysis}
\label{sec:taskcentric}

Using a data sequence framework, we have shown that we can analyse individual or ensemble trajectories of students. However, one of the benefits of a data sequence framework is that it can be used for task centric analysis, i.e. each individual task can be analysed with respect to the sequences of a group of students to study particular outcomes such as performance. In this section we introduce task centric quantities that can be extracted using a data sequence framework. In particular, we explore the differences between high and low performance in completion behaviour for individual tasks and task types.

\begin{figure*}[!t]
    \centering
    \includegraphics[width=0.8\textwidth]{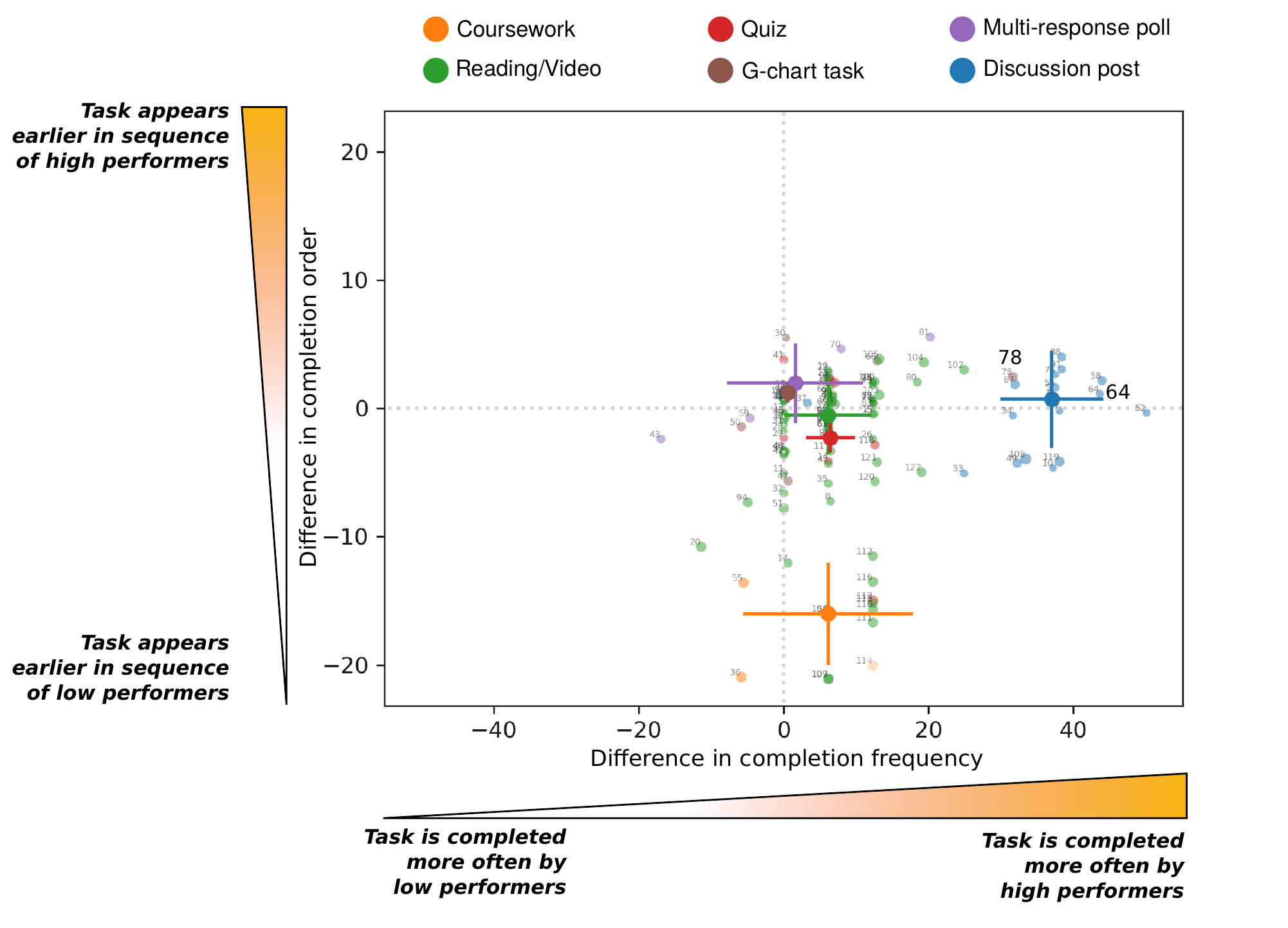}
      \caption{Comparing the high and low performing groups: The difference between the two groups in completion frequency versus the difference between the two groups in mean completion order. The tasks are identified by their task ID and colored by type (the six types are listed in the legend). For each task type, the median and inter-quartile range is plotted. If a task appears in the upper half, the task has been completed earlier on average by the group of high performers. If a task is located in the right half, the task has been completed more frequently by the group of high performers. Discussion posts, multi-response polls and G-chart tasks are completed significantly earlier and more frequently by high performers suggesting the importance of active learning. Coursework tasks appear earlier in the sequence of low performers because low performers have completed less of the course by the time the coursework is due for submission.}
    \label{fig:4}
\end{figure*}

The tasks are of six types: Coursework; Reading/Video; Quiz; G-chart task, Multi-response poll, and Discussion post. 
For each task, we calculate: (i)  the difference in the frequency of completion between high and low performers; (ii) the difference in the mean position in the completion sequence between high and low performers.
% Another task specific measure to compare between the low and high performing learners is to explore the relative frequency and the mean point in the sequence that a task was completed. 
% For example, a particular task may provide a more thorough understanding when completed after another task but not before it. 
The results for all tasks are shown in Fig.~\ref{fig:4} identified by their Task ID and colored by their task type.
%displays the difference between the frequency of completion of a task between the two groups of learners against the difference between the mean completion step of a task. 
Deviations from the (0,0) point in the centre of Fig.~\ref{fig:4} correspond to differences between low and high performers.  
If a task appears in the upper half, the task has been completed earlier on average by the group of high performing learners. If a task is located on the right half, the task has been completed more frequently by the group of high performing learners. Therefore, a task located in the upper right quadrant is completed earlier (in the learner sequence - not necessarily in time) and more frequently by high performance learners. 
%Conversely, tasks located in the lower left quadrant are completed earlier and more frequently by low performance learners. 

In Fig.~\ref{fig:4} we also show the median and interquartile range for each task type.  The median for the 'Reading/Video' tasks appears almost directly at the centre of the plot, indicating that completing a video or a reading task is not correlated with high or low performance. 
In contrast, other task types show significant deviations between the two groups. Specifically, discussion posts or interactive tasks such as multi-response submissions (usually requires a learner to make a calculation) and G-chart tasks (learners must submit a choice from a selection of answers) have a median that is located in the upper right quadrant, i.e., they are completed more frequently and earlier by high performers.
In addition, coursework tasks appear significantly earlier in the sequences of low performers. Although this might seem counterintuitive at first sight, it is consistent with the fact that low performers tend to skip more tasks but do complete courseworks.

% Other task types include the G-chart submission where a learner must choose from a selection of answers (usually just yes or no) and multi-response submissions which usually require a learner to make a few calculations and submit their values into input fields. These two task types were predominantly found in the upper right quadrant and were correlated with higher performance.

Beyond general task types, we can identify specific tasks that may be correlated with high performance. For example, tasks 64 and 78 (highlighted in Fig.~\ref{fig:4}) exhibit high frequency and early completion by the high performing learners. Task 64 was an open discussion where learners were asked to perform two calculations and then discuss the results; task 78 required learners to make a decision about picking stocks for a portfolio. Both of these tasks require the learner to understand the previous content. The public nature of the answers among their peers might put off learners that lack confidence in their answers. In the case of discussion tasks, where a learner must submit a public answer to a tutor's question, 19/20 discussion tasks are found in the right half, where high performers complete these tasks more frequently, and 15/20 are found in the upper right quadrant, where they also appear earlier in the sequence of high performers.

The bottom half contains a number of tasks such as the coursework tasks 36, 54, 55 and 115. 
%(tasks 54 and 55 constituted two different submissions of the same coursework). 
As discussed above, coursework tasks appear significantly earlier in the sequence of low performers as a reflection of the fact that they have completed fewer tasks by the time the coursework is due for submission.
Hence low performance learners skip ahead to reach the coursework. 
Other tasks completed earlier and more often by low performing students include a number of tasks from session 2 (tasks 17, 20, and 29) and session 9 (tasks 108 and 109) which appear because low performing students skipped the beginning of both sessions. %2. This is also observed for various tasks in session 9; 
%session 9 contained a coursework task (task 115) and it appears that low performing students jumped ahead to this session to learn the material for the impending coursework deadline.
%Task 88 is also an interesting task (we identified this task using the quantile regression in Fig.~\ref{fig:comparison_2groups_2}B) that was skipped by a large group of learners likely due to its demanding nature. Other tasks include 56 and 70, which required the learner to answer a multi-choice question and submit an answer to a conceptual question regarding the applicability of `P/E ratios' respectively. 
%Tasks 96-104 all appear to be completed earlier by the high performing group - we saw earlier that these tasks comprised Session 8 of the course and had been skipped by the low performing group likely in an effort to start the material for the coursework in Session 9.

%The quiz tasks don't seem to show any particular correlation. Indeed we wouldn't expect them to; the visualisation of data here does not inform us of the scores that each learner received and simply completing a quiz doesn't provide valuable information about learning. 

The results in this section, again, highlight the need for high-resolution temporal data that can differentiate between individual tasks. At the session-level we would be unable to differentiate clearly between task types. We show that the sequence data framework and the associated analyses are able to highlight tasks that are associated with anomalous engagement behavior. This provides learning designers and educators with an appropriate method for analyzing the course design and evaluating any parts that may require re-design.

\section{Conclusion}

In general, current temporal analyses of student data tend to obfuscate the longitudinal relationships between events through averaging, coarse graining or feature extraction. In this paper, we have used a data sequence framework to form the basis of a set of data-driven analyses and a Bayesian sequence model to investigate high-resolution task completion data of learners. We also show that using this framework we can introduce a task-centric analysis which is capable of informing learning design.
In general we identify aggregate and individual learner behaviors; explore the relationship between course structure and learner trajectories; reveal the connection between completion patterns and task types;  and identify points in the course where learners might benefit from intervention.

\paragraph{Summary of paper}

The first section used the data sequence model to analyze the ensemble of learner trajectories relative to the nominal task order. We highlighted the fact that although learners generally follow the nominal course structure, a number of exceptions occur, for different reasons, at different junctures in the course. We identified various behaviors such as bimodality and branching from nominal task ordering that occurred as a consequence of groups of learners taking alternative approaches to task completion. Our comparison between learner trajectories and nominal course structure revealed large deviations in different parts of the course and increasingly as the course progresses. Whilst not given here, analyzing the causal effects of such behaviors could provide the educator with an improved understanding of their course design, and about individual learners undertaking the course.

In the second section, we compared a high-resolution approach with a coarse-grained approach to temporal analysis. The transition probability between tasks and sessions was used to study the deviations when learners transition between non-sequential tasks. The analysis at the coarse-grained level of inter-session transitions revealed specific sessions in the course structure where deviations from the nominal course structure occurred, highlighting the importance of an analysis at different levels of time resolution to understand the interactions of learners with the material~\citep{Lund2017}.

In the third section, we furthered our comparison of high-resolution and coarse-grained approaches to evaluate differences between high and low performance learners. We began our exploration of the differences in completion behaviors between high and low performing learners. Having split the learners into two groups according to the median grade, we showed that low performing learners followed more strictly the task structure within a session but were less likely to follow the session structure throughout the course. The analysis also showed that low performers tend to skip a number of sessions in order to complete coursework tasks.

In the penultimate results section, we introduced a Bayesian probabilistic model to learn the trajectories of task completion sequences. Using the top 25\% and bottom 25\% groups of students (a bimodal performance distribution) we showed that the task completion sequences could be learned and used for predicting student performance.

In the final section, we showed that our data sequence model could be used for task-centric analyzes. We examined the link between the completion of individual tasks and learner performance. Our analysis showed that completion of task types related to active learning (such as  discussion posts, multi-response polls, or G-chart tasks) were correlated with above median performance. 
%Does skipping such a task result in a learner under-performing indicating a causal relationship? 
%Although no diWe do not think so, but we do believe that 
The completion of such tasks, some of which are also visible to the peer learners, might indicate a level of understanding and confidence in the knowledge of the material around that task. An educator may consider encouraging the completion of some of those key tasks (or making them compulsory) in order to encourage the learner to learn the material in order to complete that task. 
%Alternatively, the learner could be nudged into completing that task with prompts. An example could be: `Students that complete this task are more likely to attain a higher grade'.

We found a clear distinction between high and low performers as regards to the types of tasks, how frequently they were completed, and at what point in their sequence. These findings could be considered in the context of Bloom's Taxonomy of Learning \citep{Bloom1956}. For example, discussion-based tasks require skills toward the higher end of the taxonomy, whereas tasks that comprise reading texts and watching videos require skills towards the lower end. It is then not surprising that tasks in the former category were more commonly completed and earlier in the study sequence of high performing students. Identifying these types of tasks could help tutors structure the courses to reinforce the importance and build-up towards such tasks, as well as identifying differences between groups of learners. 
%restructure or make compulsory these critical tasks and (ii) help tutors identify or differentiate between types of learners.
These results exemplify through a small illustrative example the power of relatively simple sequence analysis methods applied to learning data, and the potential for their use on available data to reveal the connections between learning patterns, performance, and course design.

\paragraph{Contributions}

This paper has provided: (1) methodological contributions (data-driven metrics and sequence modelling); (2) insights into the resolution of temporal data; (3) an evidence base to produce an actionable software package within our LMS.

The methodological contributions of this paper are three fold: (i) We show that a data sequence framework provides an appropriate format for investigating the temporal relationships of learning data; (ii) We introduced new data-driven analytical metrics for investigating sequential data; (iii) Using a Bayesian sequence model we are able to learn the trajectories of learners and show that it can be used for predicting performance.

The research in this paper has also provided necessary contributions within our Business School. We are currently developing a software module that incorporates the data sequence model into our LMS for two main purposes: (i) To identify anomalous student engagement behaviours; (ii) To evaluate the course design and identify areas that may require re-design.

\paragraph{Future Work}

Whilst the majority of on-line courses have a designed linear structure, it is clear that cognitive development does not progress through a fixed sequence of events. The alternative is to present each course component in parallel, without a displayed intended structure, and students can begin at any point and transition between any components of the course. In such a system, our Bayesian sequence model could still be applied given that it is able to characterise the complete state-space of possible trajectories and isn't necessarily dependent on the ground truth course structure. Given this, we intend to pursue the Bayesian model further.

A second direction would be to look at higher order network models which don't make Markovian assumptions. Considering the full trajectories of each student could yield higher order temporal dependencies that may better model the transitions between tasks.

\section*{Declaration of Conflicting Interest} % The \section*{} command stops section numbering
The author(s) declared no potential conflicts of interest with respect to the research, authorship, and/or publication of this article.

\bibliographystyle{apalike} 
\bibliography{refs}

\end{document}